\documentclass[useAMS,usenatbib]{mnras}
\usepackage{amsmath}
\usepackage{amssymb}
\usepackage{graphicx}
\usepackage{appendix}
\usepackage{color}
\usepackage{multirow}
\usepackage{rotating}
\usepackage{wrapfig}
\usepackage{natbib}
\newcommand{\lSect}[1]{{\label{sec:#1}}}
\newcommand{\lFig}[1]{{\label{fig:#1}}}

\newcommand{\FIGFF}[2]{{\ref{fig:#2}{#1}}}

\newcommand{\FIG}[2]{{Fig.~\FIGFF{#1}{#2}}}
\newcommand{\Fig}[1]{{\FIG{}{#1}}}

\newcommand{\Figure}[1]{{Figure~\FIGFF{}{#1}}}

\newcommand{\Sectff}[1]{{\ref{sec:#1}}}
\newcommand{\Sect}[1]{{\S~\Sectff{#1}}}

\newcommand{\Msun}{\ensuremath{\mathrm{M}_\odot}}
\newcommand{\Rsun}{\ensuremath{\mathrm{R}_\odot}}

\newcommand{\KEPLER}{\texttt{KEPLER}}
\def\lesssim{\mathrel{\hbox{\rlap{\hbox{\lower5pt\hbox{$\sim$}}}\hbox{$<$}}}}
\def\gtrsim{\mathrel{\hbox{\rlap{\hbox{\lower5pt\hbox{$\sim$}}}\hbox{$>$}}}}
\def\gtaprx {\lower .1ex\hbox{\rlap{\raise .6ex\hbox{\hskip .3ex
	{\ifmmode{\scriptscriptstyle >}\else
		{$\scriptscriptstyle >$}\fi}}}
	\kern -.4ex{\ifmmode{\scriptscriptstyle \sim}\else
		{$\scriptscriptstyle\sim$}\fi}}}
\def\ltaprx {\lower .1ex\hbox{\rlap{\raise .6ex\hbox{\hskip .3ex
	{\ifmmode{\scriptscriptstyle <}\else
		{$\scriptscriptstyle <$}\fi}}}
	\kern -.4ex{\ifmmode{\scriptscriptstyle \sim}\else
		{$\scriptscriptstyle\sim$}\fi}}}

\title[Magnetar Powered Ordinary Type IIP Supernovae]{Magnetar Powered Ordinary Type IIP Supernovae}

\author[Sukhbold \& Thompson]{Tuguldur Sukhbold \& Todd A. Thompson \\
Department of Astronomy and Center for Cosmology \& Astro-Particle Physics, The Ohio State University, Columbus, Ohio 43210}

\begin{document}

\maketitle
\label{firstpage}
\begin{abstract}
We investigate the properties of Type IIP supernovae that are 
dominantly powered by the rotational kinetic energy of the 
newly born neutron star. While the spin-down of a magnetar 
has previously been proposed as a viable energy source in the 
context of super-luminous supernovae, we show that a similar  
mechanism could produce both normal and peculiar Type IIP 
supernova light curves from red supergiant progenitors for a 
range of initial spin periods and equivalent dipole magnetic 
field strengths. Although the formation channel for such magnetars 
in a typical red supergiant progenitor is unknown, it is 
tantalizing that this proof of concept model is capable of 
producing ordinary Type IIP lightcurve properties, perhaps 
implying that rotation rate and magnetic field strength may 
play important roles in some ordinary looking Type IIP 
supernova explosions.
\end{abstract}


\begin{keywords}
magnetars, supernovae
\end{keywords}

\section{Introduction}
\lSect{intro}

The most common massive star supernovae in the universe, by 
number, are the explosions of red-supergiant (RSG) progenitors 
with initial masses of $\sim$ 9-18 \Msun\ \citep{Sma15}, and 
with kinetic energies on the order of $\sim10^{51}$ ergs 
\citep{Pej15}. These explosions create light curves of Type IIP, 
which gradualy releases the shock deposited energy to produce 
roughly a constant luminosity for about 3 months 
\citep[e.g.,][]{Fil97,Arc12}. The most intensely studied 
mechanism for these explosions is the delayed-neutrino driven 
mechanism \citep{Col66,Arn66,Bet85}. While the neutrino 
mechanism is rather successful for the low energy explosions of 
lighter progenitors, there is no consensus yet in the supernova 
community on more massive progenitors that would provide the 
observed energies of typical supernovae 
\citep[][and references therein]{Jan16}. Inspired by 
\citet{Bod74} and the application of neutron star models to 
luminous supernovae by \citet{Woo10} and \citet{Kas10}, in this 
work we explore the scenario where the rotational energy source 
plays a dominant role throughout the explosion of a RSG 
progenitor.
 
Soon after the discovery of pulsars, it was suggested that an 
embedded pulsar might power the light curves and explosions of 
ordinary supernovae \citep{Ost71}. More recently, the idea has 
gained new traction as a possible explanation for gamma-ray 
bursts \citep[e.g.,][]{Uso92,Tho04,Uze06,Buc09,Met11} and other 
unusal supernovae \citep{Aki03,Mae07}. The pulsars required for 
these transients have atypical properties in that their field 
strengths and rotation periods are extreme. Today, such models 
are thus commonly referred to as ``magnetar'' models rather 
than ``pulsar'' models.

Following \cite{Mae07}, \citet{Woo10} and \citet{Kas10} 
independently promoted the idea that magnetars might 
underlie the production of a broad class of hydrogen poor 
super-luminous supernovae (SLSN-I), which are brighter than 
10$^{43}$ ergs s$^{-1}$ for a time longer than typical Type 
Ia supernovae (a couple of weeks); see \citet{Qui11,Gal12} and 
\citet{Nic15} for recent reviews of SLSN-I. Since 2010, many 
studies have interpreted the light curves of SLSN-I as the 
product of magnetar spin down \citep[e.g.,][]{Ben14,Nic13,Met15,Don16}.

Consider a scenario where a weak explosion is initiated in a 
RSG progenitor, and soon afterwards a magnetar, which formed 
from the collapse of a spinning progenitor core, starts to 
deposit energy. The exact nature of the formation is not well 
understood, an issue we revisit in \Sect{conclude}. In general, 
such a scenario would be representative of a case where the 
original neutrino wind becomes increasingly magnetically 
dominated as the proto-neutron star cools 
\citep{Tho04,Met07,Met11}, eventually transitioning to a 
highly relativistic ``pulsar''-type wind, depositing the pulsar 
rotational kinetic energy on the spin-down timescale.

Since the brightness and duration of the light curve plateau 
phase scales with the promptly deposited explosion energy as 
$\rm\propto E^{5/6}$ and $\rm\propto E^{-1/6}$ respectively 
\citep{Pop93}, the weak explosion alone, without any magnetar 
input, will create a long and dim transient compared to a 
typical Type IIP case. For the magnetar to transform this weak 
explosion into a typical one, it will need to inject a 
significant amount of energy in a short timescale, so that the 
plateau phase is brighter and transitions to the nebular phase 
sooner.

Approximating the initial rotational kinetic energy of the 
magnetar as $\rm E_m\approx2\times10^{52}P^{-2}_{ms,i}$ ergs, 
where $\rm P_{ms,i}$ is the initial spin period in milliseconds, 
we can see that if it is to yield a final kinetic energy of an 
ordinary Type IIP explosion roughly between 0.5$-$2$\times10^
{51}$ ergs, the initial spin needs to be approximately between 
3 and 6 ms. At late times (after the plateau phase) the tail of 
the light curve must be dominantly powered by the radioactivity, 
$\rm L_m(t_{late}) < L_{Co}(t_{late})$, where $\rm L_m$ is the 
spin down luminosity and $\rm L_{Co}$ is the luminosity from the 
decay of $^{56}$Co, resulting from a typical Type IIP yield of 
$^{56}$Ni. For vacuum dipole spin-down the magnetar luminosity 
is $\rm L_m\approx10^{49}B_{15}^2P_{ms}^{-4}\ ergs\ s^{-1}$, 
where $\rm B_{15}$ is the magnetic field strength in $10^{15}$ G 
and an angle of $\pi/6$ was assumed between the rotational and 
magnetic axes. Taking $\rm t_{late}$ as 150 days and adopting a 
typical $^{56}$Ni mass of 0.1 \Msun, we see that the spin down 
timescale, $\rm t_m = E_m/L_m$, needs to be shorter than roughly 
a day for $3\rm<P_{i}<6$ ms, or the constant dipole magnetic 
field needs to be larger than roughly $10^{15}$ G. At the other 
end, invoking more extreme conditions with $\rm P_{i}\sim 1$ ms 
and field strengths of $>10^{16}$ G may end up producing a 
typical IIP-like light curves in some situations, but will 
ultimately yield much higher energies and also may overproduce 
$^{56}$Ni.

These basic considerations, though based on an idealized 
situation where the magnetar keeps indefinitely injecting energy, 
that is efficiently thermalized, based on vacuum dipole emission 
(braking index of $\rm n=3$) and a constant magnetic field 
strength, demonstrate that the rotational rates and magnetic 
field strengths of such IIP-powering magnetars must be larger 
than what is commonly inferred from pulsars. In this work, 
through a set of numerical experiments we explore the question 
of over what parameter space the classical magnetar spin down 
scenario can transform a weak explosion model of a RSG into one 
where the light curve characteristics, $^{56}$Ni yields, and 
final kinetic energies are close to those of typical IIP 
supernovae.


\section{Numerical Calculations with \KEPLER}
\lSect{numeric}

We calculate a set of magnetar powered RSG explosion models 
using the 1D implicit hydrodynamic code \KEPLER\ \citep{Wea78}. 
All calculations start with a RSG progenitor model from 
\citet{Suk14} with an initial mass of 15 \Msun. At the time of 
core collapse this model had a radius of 841 \Rsun\ and a total 
mass of 12.6 \Msun, of which the outermost 8.3 \Msun\ was in 
the H-rich envelope. We first launch a weak explosion by using 
the moving inner boundary method, i.e. ``piston-scheme'' 
\citep{Woo95,Suk16a}, so that the 
final kinetic energy of the ejecta is only about $\sim 5 \times 
10^{49}$ ergs. This explosion synthesized roughly 0.16 \Msun\ 
of $^{56}$Ni, but due to late time fallback only about $\sim 
0.015$ \Msun\ is ejected. The synthesized $^{56}$Ni mass is on the 
large side because for this demonstration model the piston was 
deliberately placed at the edge of the iron core (1.43 \Msun), 
which is significantly deeper than the mass cut that could 
represent a fully neutrino-driven explosion ($\sim$1.6 \Msun\ 
for the same model in \citet{Suk16a}). This choice primarily 
stems from our expectation that magnetar input, with our current 
description, would not significantly contribute to the $^{56}$Ni 
synthesis, which we discuss further in \Sect{conclude}. 
As in \citet{Woo10}, we do not specify the physical nature of 
the explosion initiation. 

\begin{figure}
\centering
\includegraphics[width=.48\textwidth]{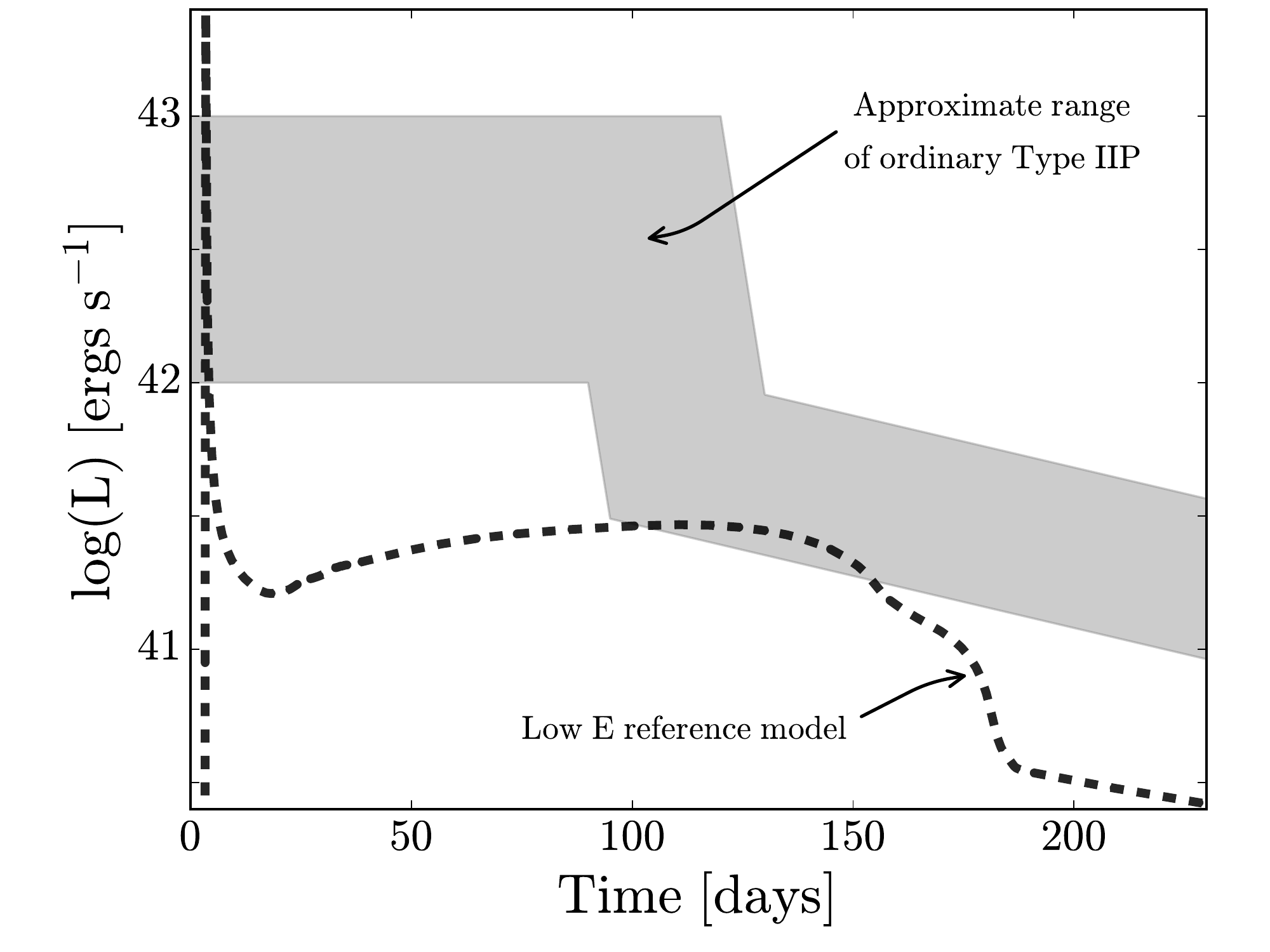}
\caption{The light curve from the low energy $5\times10^{49}$ ergs 
explosion (dashed black), without any further energy 
input, is shown in comparison with an approximate 
luminosity band of typical Type IIP supernovae. With 
much lower energy, the resulting light curve is much 
dimmer and has a long lasting plateau phase. \lFig{lowE}}
\end{figure}

As expected, this low energy explosion, without any 
additional energy input produces a long lasting dim transient (\Fig{lowE}). 
Taking the scaling relations based on the survey of model Type 
IIP light curves from \citet[][Eqs. 15 and 17]{Suk16a}, one gets 
a plateau luminosity of $\rm L_{p} = 2.3 \times10^{41}\ ergs\ 
s^{-1}$ and a duration of the plateau, including the effects of 
radioactivity, is $\rm \tau_p = 176$ days. These values are in a 
good agreement with the low energy explosion model shown in 
\Fig{lowE} as a dashed black curve. The plot also shows a rough 
luminosity range that represents the typical IIP light curves 
(light gray band): $\rm 10^{42}<L_p<10^{43}\ ergs\ s^{-1}$, $\rm 
90<\tau_p<130$ days and the tail representing $^{56}$Co decay 
luminosity for $\rm 0.05<M_{Ni}<0.2$ \Msun\ \citep{Pej15}. Note 
that the ``normal'' light curve band corresponds to a much brighter 
and briefer plateaus than in the low energy explosion 
reference model (dashed curve in \Fig{lowE}).

\begin{figure*}
\centering
\begin{tabular}{cc}
\includegraphics[width=.48\textwidth]{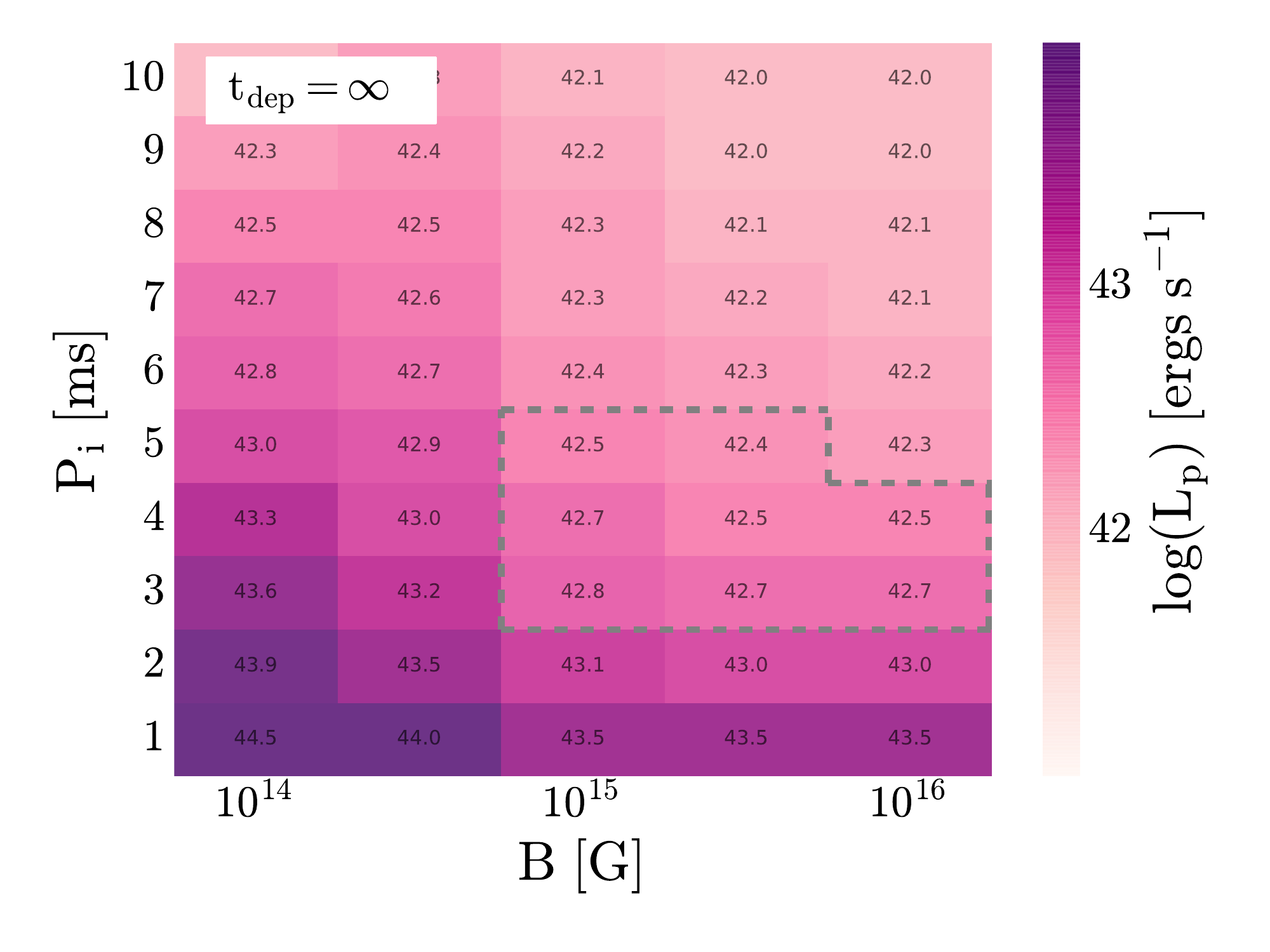} &
\includegraphics[width=.48\textwidth]{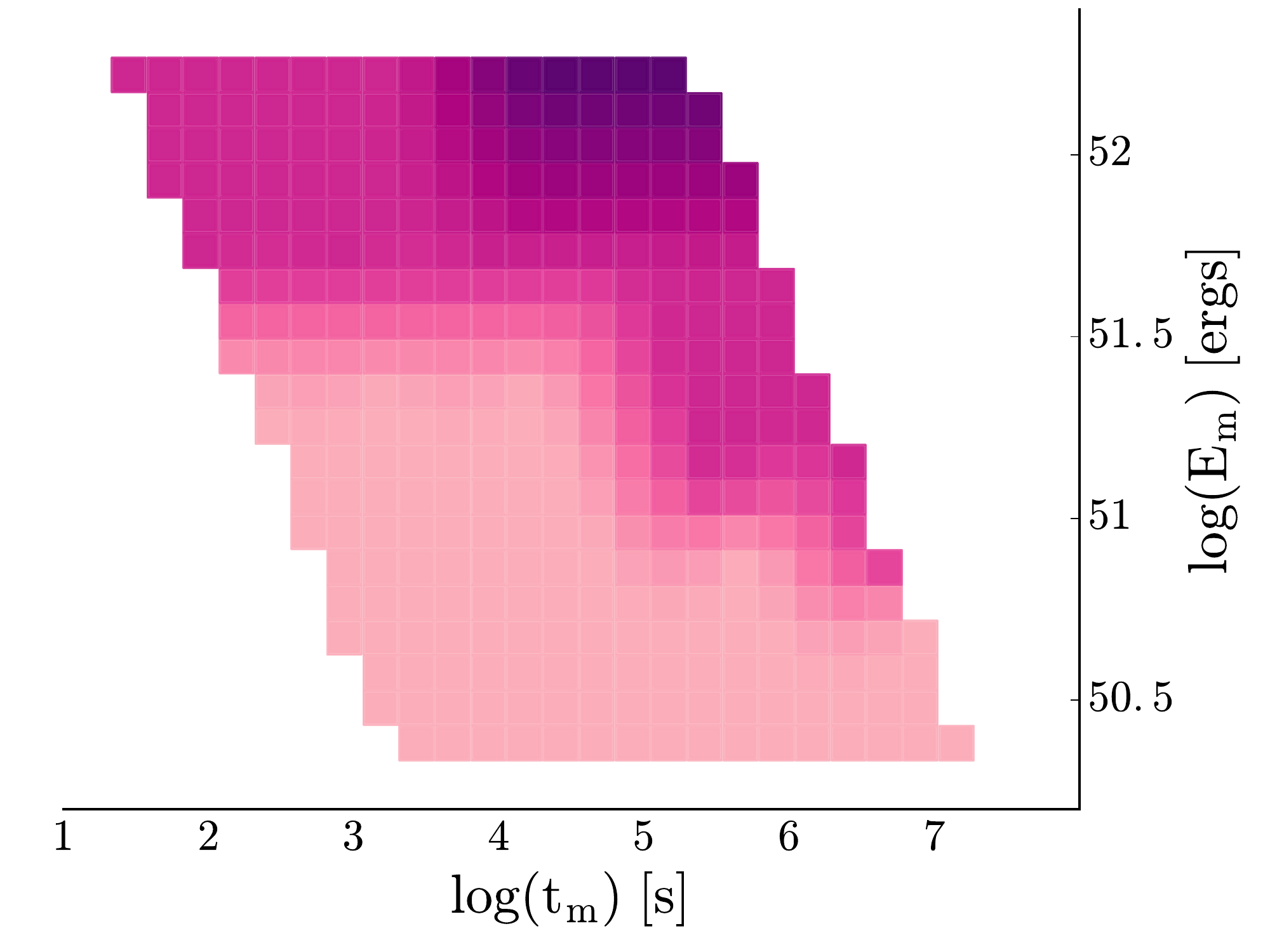} \\
\includegraphics[width=.48\textwidth]{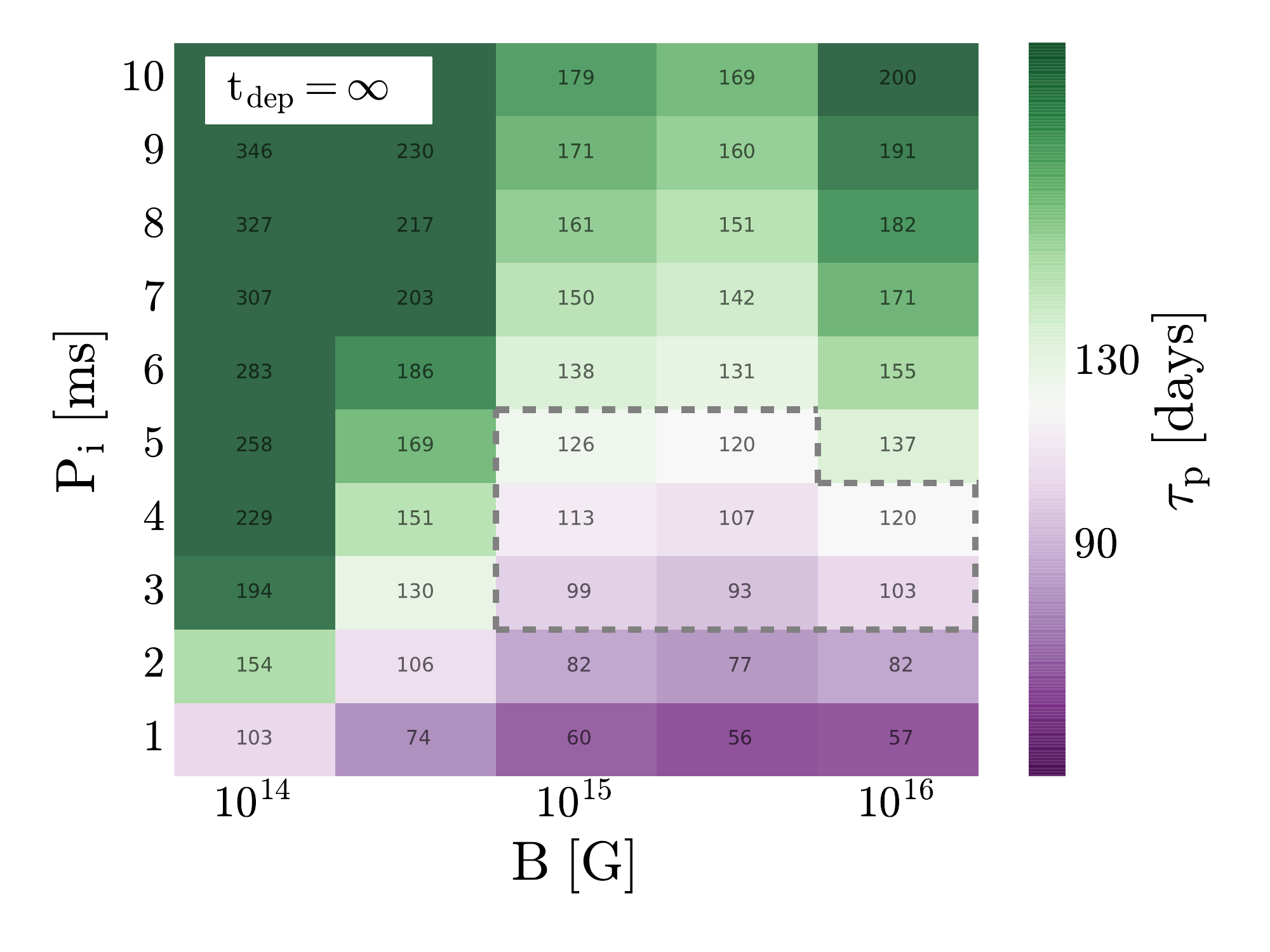} & 
\includegraphics[width=.48\textwidth]{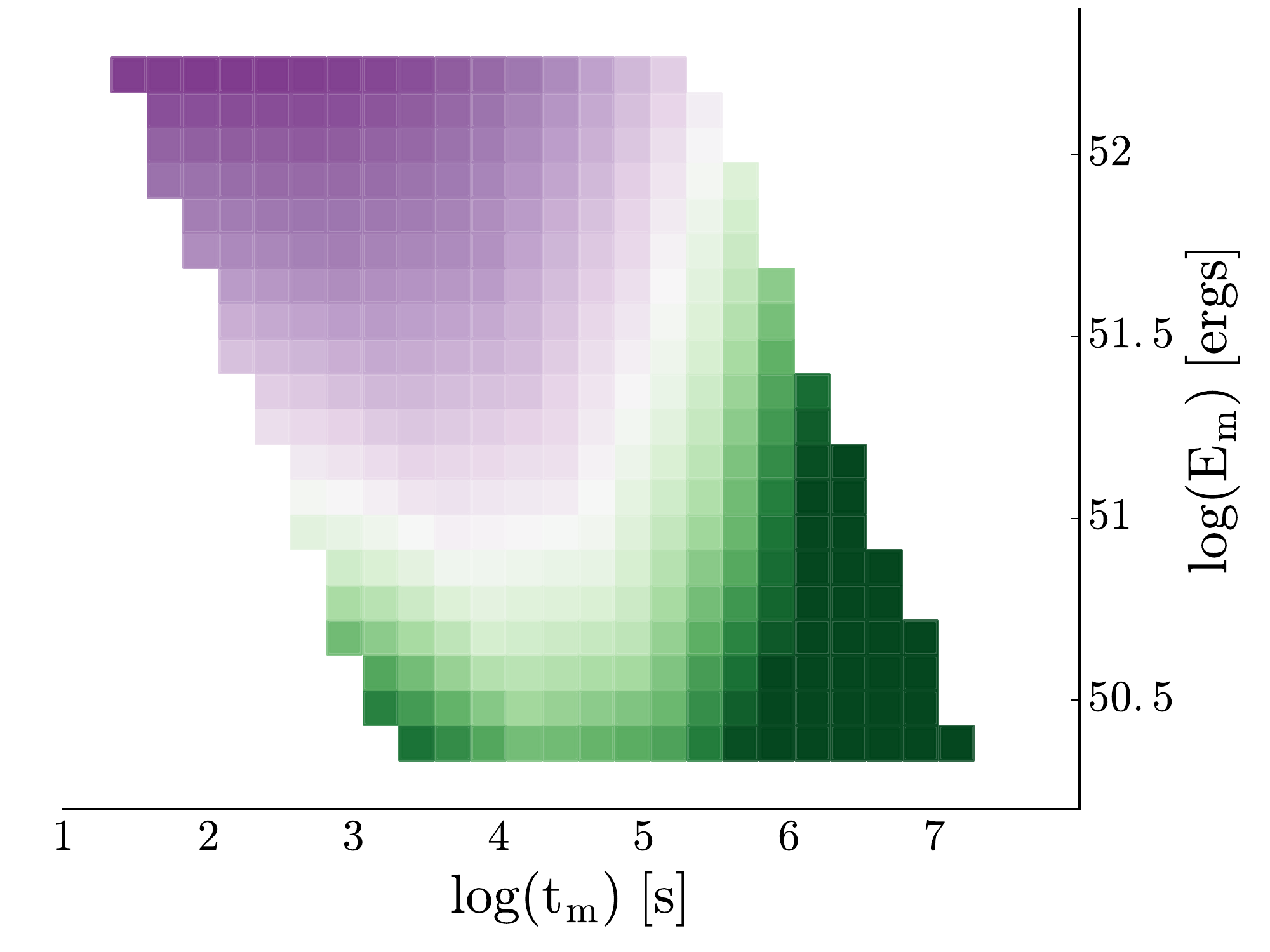}
\end{tabular}
\caption{Heatmaps of lightcurve plateau luminosity (top) and 
durations (bottom) for magnetar powered light curves as a function 
of the magnetic field strength B and the initial spin $\rm P_i$ 
(left), and as a function of initial spin-down timescale $\rm t_m$ 
and rotational kinetic energy $\rm E_m$ (right, interpolated).
Each case starts with low energy ($5\times10^{49}$ ergs) piston 
explosion of a $\rm M_{ZAMS}$=15 \Msun\ RSG progenitor, and the 
magnetar energy was deposited from 0.25 seconds until the end of 
calculation. The plateau duration, $\rm \tau_p$, is measured as 
the time from the explosion until when the photospheric 
radius dips below $10^{14}$ cm, and the plateau luminosity is 
measured as $\rm L_p=L(\tau_p/2)$, see text for details. Regions 
surrounded by dashed lines in each panel denote the ranges of 
initial magnetar spin and magnetic field strength that produces 
typical Type IIP light curves with $\rm 90<\tau_p<130$ days and 
$\rm 42<log(L_p)<43\ ergs\ s^{-1}$. \lFig{tdepinf}}
\end{figure*}

The piston that drives the low energy explosion first moves 
inward from its initial location at the edge of the iron core 
to a minimum radius of $10^7$ cm in 0.25 seconds. Once it starts 
to move outwards, we begin to deposit energy to the inner part of 
the ejecta as heat according to the vacuum dipole spin-down 
formulation, $\rm L_m(t)=E_m t_m^{-1} (1+t/t_m)^{-2}\ ergs\ 
s^{-1}$. The evolution of the ejecta is followed for $\sim350$ 
days including the energy from radioactivity, and the light 
curves are approximately calculated with flux-limited diffusion. 
The magnetar parameters were varied from B=$10^{14-16}$ G for 
the constant dipole magnetic field strength at the equator, and 
between 1 and 10 ms for the initial rotational period. These 
spins correspond to a range of initial rotational kinetic 
energies between $0.2-20\times10^{51}$ ergs, and spin-down 
timescales between 20 seconds and 230 days.

Once the explosion is initiated, we expect a neutrino-driven wind 
to emerge from the cooling proto-neutron star 
\citep{Dun86,Jan96,Bur95}. For a given magnetic field strength 
and spin period, the flow will become increasingly relativistic 
over the Kelvin-Helmholz cooling timescale as the Alfv\'{e}n 
point in the flow approaches the light cylinder 
\citep{Tho04,Met07,Met11}. The energy injection rate at these 
very early times is higher than implied by the vacuum dipole 
expression for the same $\rm B$ and $\rm P$, and may be 
complicated as the neutron star contracts and its convection 
changes character throughout the cooling epoch \citep{Pon99,Rob12}. 
However, because the long-term spindown behavior more directly 
affects the eventual lightcurve shape and dynamics than the 
detailed evolution at these very early times, we simply assume 
the vacuum dipole energy injection formula. For these 
purposes $\rm B$ and $\rm P$ should be interpreted as defining 
the rotation energy reservoir and the reference energy loss rate.

The effect of late time magnetar powered light curves have been 
extensively studied in the context of SLSN-I emerging from 
stripped cores \citep[e.g.,][]{Ins13}. In general, 
with shorter spin down timescale, most of the magnetar energy 
input is lost to adiabatic expansion, while with longer timescale 
more energy is channeled into radiation and produces luminous 
light curves. The same generic behavior is also seen in 
our calculations, however, with the extended envelope 
of the RSG progenitor the light curves present a diverse structure, 
and with a larger ejecta mass the magnetar powered Type II 
light curves are generally less luminous than Type I cases.

\Figure{tdepinf} shows the resulting light curve plateau properties 
from magnetar powered explosions when the energy was injected 
until the end of calculation ($\sim$350 days). The plateau duration, 
$\rm \tau_p$, was conservatively measured as the time span from 
the beginninng of explosion until when the photospheric radius 
dips below $10^{14}$ cm, while the plateau luminosity was 
measured in the middle of the plateau as $\rm L_p=L(\tau_p/2)$. 
These conditions apply well to all calculations, except those with 
the weakest field stregths and slowest initial spins. In those 
models, the spin-down timescale is comparable to the calculation 
time and in a few cases the photospheric radius does not reach 
its maximum within 300 days, resulting in very long $\rm \tau_p$.

For a given initial spin, the most luminous light curves emerge 
from models with weakest field strengths since the spin-down 
timescale is longest. For a given field strength, the spin-down 
timescale decreases with faster initial spin, but due to the 
increasing energy budget, much more energy is channeled into 
radiation compared to a slower spin model. The most luminous 
model with $10^{14}$ G and initial spin of $\rm P_{ms}$=1 
exceeds a peak luminosity of a few times $\rm 10^{44}\ ergs\ 
s^{-1}$. In models with stronger fields the plateau luminosity 
never exceeds of $\rm 10^{43}\ ergs\ s^{-1}$, except the few 
with the fastest initial spins. From the other side, the least 
luminous models come from strongest fields and lowest energy 
budgets. In all of these calculations, the deposited energy is 
much larger than the initial weak explosion energy of 
$5\times10^{49}$ ergs, and thus all the models are 
significantly more luminous than the reference light curve shown 
in \Fig{lowE}.

The behaviour of the plateau duration is slightly more 
complicated. With more energy deposited promptly (i.e. 
shorter spin-down timescale), the ejecta will be strongly ionized 
and will expand faster, resulting in brighter and briefer 
plateaus. While with more gradual deposition, the magnetar 
energy extends the plateau phase by supporting ionization, in 
much the same way as done in through radioactive decay energy 
\citep[see][]{Kas09}. This is why the plateau duration increases 
with slower initial spin, for a given field strength. However, 
note that for a given initial spin the plateau duration first 
shortens with increasing field strength, and then starts to 
lengthen again for $\rm B>10^{15}$ G. For smaller field strengths, 
the spin-down luminosity is always greater than or comparable 
to the luminosity from radioactive decay, while at higher field 
strengths it is always much weaker than the decay luminosity 
at late times. 

Consistent with the results found in \citet{Suw15}, none of our 
models produced extra $^{56}$Ni in addition to the initial 
explosion. However, in all of the models the fallback that occurs 
in the original reference model does not occur and thus they all 
receive power from the decay of 0.16 \Msun\ $^{56}$Ni. This has 
little relevance for the plateau phase of the lightcurve when the 
magnetar is slowly depositing energy, but it can significantly 
extend the plateau phase when nearly all of magnetar energy is 
deposited promptly. If the energy contribution from radioactivity is 
removed, the plateau duration monotonically decreases with 
increasing field strength as expected.

The regions bounded by a dashed line in \Fig{tdepinf} highlight 
the models that have similar plateau durations and luminosities 
to typical Type IIP supernovae. Note the plateau luminosities 
are between $10^{42}$ and $\rm 10^{43}\ ergs\ s^{-1}$ for most 
models with $\rm P_{i}>3$ ms, and so this region is primarily 
bounded by the plateau durations, except when the spin-down 
timescale is long with weaker field strengths and slower initial 
spins. As expected (\Sect{intro}), this region covers mostly 
models that were powered by magnetars with $\rm P_i$ roughly 
between 3 and 6 ms, and $\rm B$ stronger than $10^{15}$ G. For 
stronger field strengths there is a slight preference for a 
slower initial spin, since the radioactive extension of the 
plateau becomes less relevant with increasing rotational energy 
of the magnetar. 

\begin{figure}
\centering
\includegraphics[width=.48\textwidth]{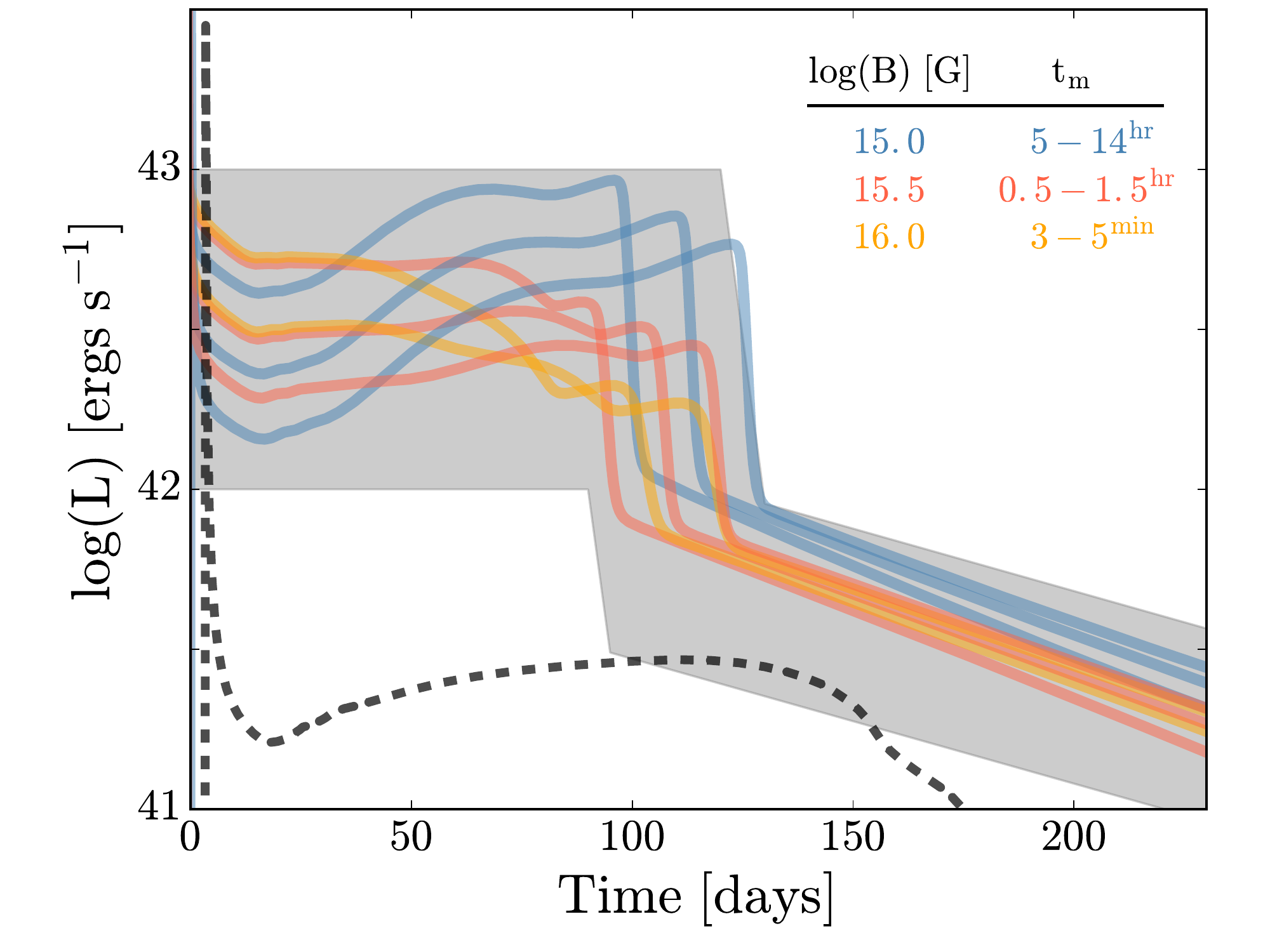}
\caption{Magnetar powered light curves (colored curves) from the highlighted 
region in \Fig{tdepinf}, where the models have plateau luminosities and durations 
that lie within the range of  observed ordinary Type IIP supernovae (gray band 
from \Fig{lowE}). The light curves from magnetars with log(B)=15 G are shown 
in blue, log(B)=15.5 G shown in red, and with log(B)=16 G are shown in orange. 
For a given field strength (curves from a given color), with increasing initial rotational energy (smaller $\rm P_i$) the light curve plateau phase is brighter and shorter. For a given initial spin, with longer the spin-down timescale (smaller B) the plateau phase brightens in time due to persistent magnetar luminosity at late times, while with shorter spin-down timescale the plateau is roughly constant or dimming with time. Observed Type IIP light curves with 
rising plateaus are often associated with blue supergiant progenitors, but 
here we show that late time magnetar energy deposition can result in such light 
curves from a normal RSG progenitor.  \lFig{lcs}}
\end{figure}

\Figure{lcs} shows light curves (colored curves) from the models that 
lie in the region bounded by dashed lines in \Fig{tdepinf}. As 
in \Fig{lowE}, it also shows the reference weak explosion model 
without any magnetar input (dashed curve), and the approximate 
luminosity band (gray band) for ``typical'' Type IIP light curves. 
Note that while some light curves have a fairly constant luminosity 
during the plateau phase (red), others are sharply increasing (blue) 
or decreasing (orange) until before they transition to nebular phase. 
The increasing luminosity during the plateau phase results when the 
magnetar deposition dominates for a time comparable or longer than 
the effective diffusion timescale of the ejecta, so that without much 
prompt energy deposition it is dimmer in the first few weeks and it 
gradually brightens due to the magnetar luminosity at later times. 
Also note that the tail is slightly brighter than models with shorter 
spin-down due to magnetar input, even though all of the models 
receive the same amount of radioactive energy. In general, this 
effect is even more prominent for weaker field strengths than 
$10^{15}$ G, where the plateau keeps rising for hundreds of days and 
several orders of magnitude in luminosity, but the resulting light 
curves are well beyond what we consider as typical of Type IIP. With 
shorter spin-down times, the magnetar efficiently deposits energy 
promptly and therefore the early light curve appears brighter, but 
it stays roughly constant or keeps dimming afterwards as the ejecta 
expands much faster without experiencing much magnetar energy input 
at later times.

These model light curves are consistent with the diversity seen in 
the observations \citep[e.g.,][]{Arc12}. The most notable are those 
with increasing plateau luminosity, which are often classified as 
peculiar and associated with a blue supergiant progenitor 
\citep[e.g.,][]{Tad16} due to their similarity with SN1987A. With 
the explosion of a compact progenitor, the light curve is initially 
less luminous due to the energy lost in adiabatic expansion. 
However, our calculations demonstrate that such light curves can 
emerge from RSG progenitors when their explosions are powered by 
magnetars. In this case, however, the progenitor is already 
extended to begin with, and the light curve starts from a lower 
luminosity because of weaker recombination of the envelope and 
gradually brightens when the magnetar keeps depositing significant 
amount of energy at late times.


\section{Discussion} 
\lSect{conclude}

We have explored the light curves emerging from an explosion of 
a RSG progenitor that was initiated by a weak piston and followed 
by energy deposition from a magnetar. Through a set of spherically 
symmetric hydrodynamical calculations with approximate radiation 
transport, we have shown that for a narrow range of magnetar 
parameters, the resulting light curves resemble what we observe 
in ordinary Type IIP supernova explosions. For the progenitor 
explored, when the initial spin of the magnetar is between 3 and 
5 ms, and its dipole magnetic field is strength greater than 
$10^{15}$ G, the energy deposition through vacuum dipole spin-down 
transforms the long-lasting dim transient, produced by the weak 
piston-driven explosion alone, into transients with plateau 
luminosities of  $\rm 10^{42-43}\ ergs\ s^{-1}$ and durations of 
90-130 days. The final kinetic energies of the ejecta are in the 
expected range ($\sim0.8-2\times10^{51}$ ergs) and since the 
magnetar deposition does not produce any extra $^{56}$Ni, the 
magnetar energy input with a short spin-down time results in 
radioactively powered light curve tails seen in ordinary Type IIP 
supernovae, as long as enough is synthesized through the initial 
weak explosion

The above-mentioned preferred range for the magnetar spin period
and magnetic field strength correspond in part to the some of the employed 
assumptions. For example, if the initial prompt energy assumed is higher, 
then lower initial spin rates would yield typical Type IIP energies at late times. 
When we repeat the calculation using a RSG progenitor with an initial mass of 12 
\Msun, and with a prompt explosion energy of $8\times10^{49}$ ergs, 
the corresponding range of $\rm P_i$ that produces typical Type 
IIP light curves shifts to $\sim4-7$ ms, because the prompt 
non-magnetar component of the energy budget for the explosion is 
larger than in the 15 \Msun\ progenitor shown in \Fig{lowE}. But 
as long as this prompt explosion energy is much smaller than the 
rotational kinetic energy of the magnetar, the same generic 
results will hold - initial period of a few ms 
($\rm P_i\sim (20/(1-E_{prompt}/10^{51}))^{1/2}$ ms) and a strong 
field strength of $\rm B>10^{15}$ G for a relatively fast initial 
spin-down time. The dependence on the progenitor 
structure is rather small  since for solar metallicity the ejecta masses and the envelope masses are 
fairly similar for progenitors with initial masses between 
$10-25$ \Msun \citep[e.g.,][]{Suk14}, which are responsible for 
majority of Type IIP supernovae \citep{Sma15}, 

Another possibility is that many of the neutron stars born in Type 
IIP supernovae start with high, but short-lived, magnetic fields 
generated by a dynamo mechanism as the neutron star cools and 
convects (e.g., \citealt{Dun92}). To probe this scenario, we have 
re-calculated all of the models but depositing energy only during 
the initial 50 seconds (\Fig{tdep50}). The amount of energy 
deposited during the initial time t is approximately 
$\rm \Delta E=E_{m,i}t(t+t_m)^{-1}$. When the spin-down timescale 
is much larger the deposition time it approaches $\rm L_mt$. 
Therefore for $\rm B < 10^{15}$ G and $\rm P_i > 4$ ms the energy 
deposition for only 50 seconds does not result in any noticable 
effect on the light curve. Accordingly the range of B and $\rm P_i$ 
that results in ordinary Type IIP plateau characteristics shifts 
to $\sim1$ ms and $\sim10^{16}$ G range. In general, the magnetar 
spin-down luminosity during the initial 50 seconds is highly 
uncertain, since the neutron star is convective and its wind 
rapidly evolves \citep{Tho04,Met11}. Therefore the magnetic field 
strengths in \Fig{tdep50} should be thought of as a proxy for the 
time-averaged magnetar luminosity during the initial 50 seconds. 
Future models should explore how pulsar driven shells might 
synthesize of $^{56}$Ni and power the shockwave, as it moves though 
dense inner core of the ejecta.

\begin{figure}
\centering
\includegraphics[width=.48\textwidth]{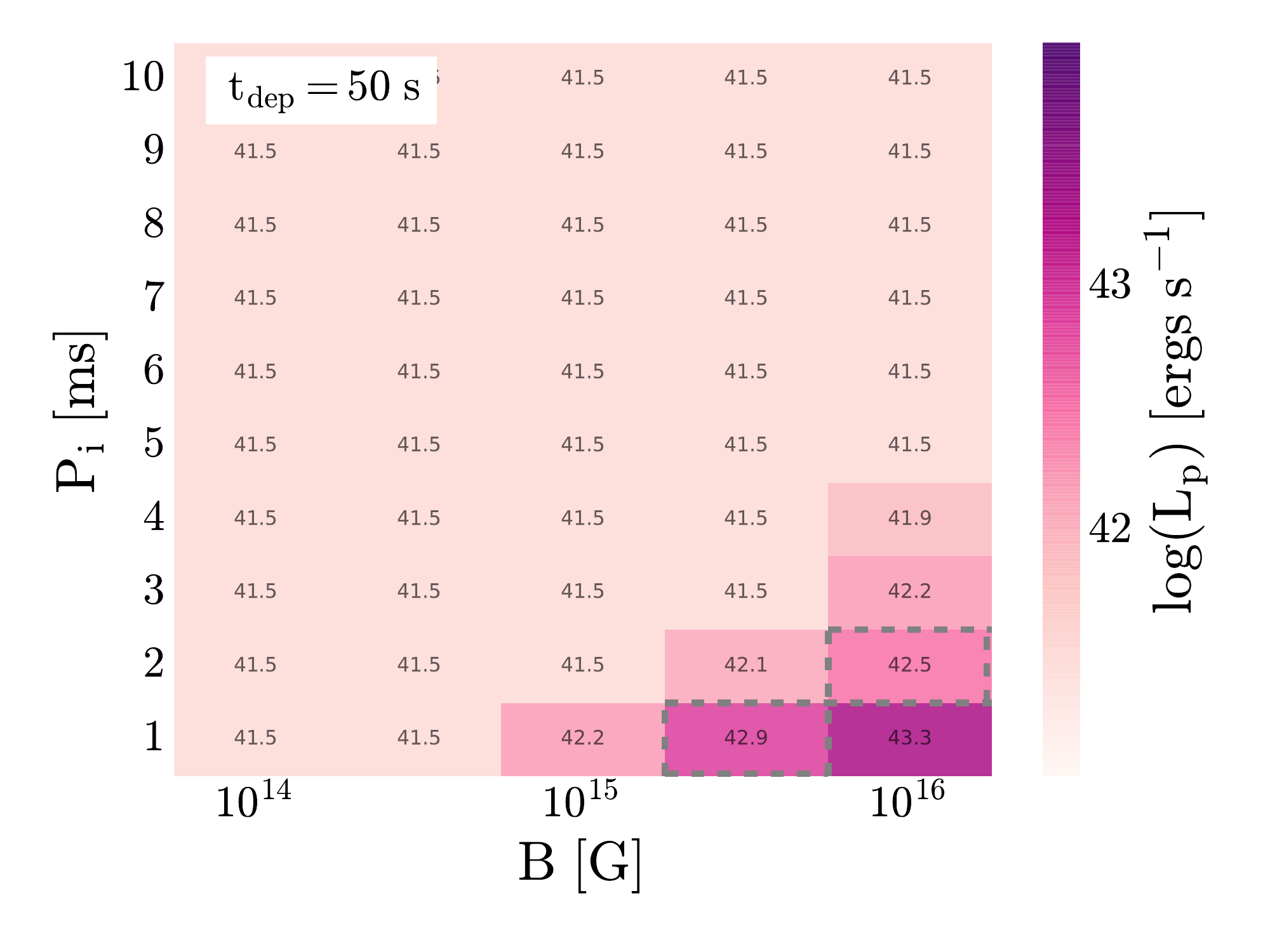}
\includegraphics[width=.48\textwidth]{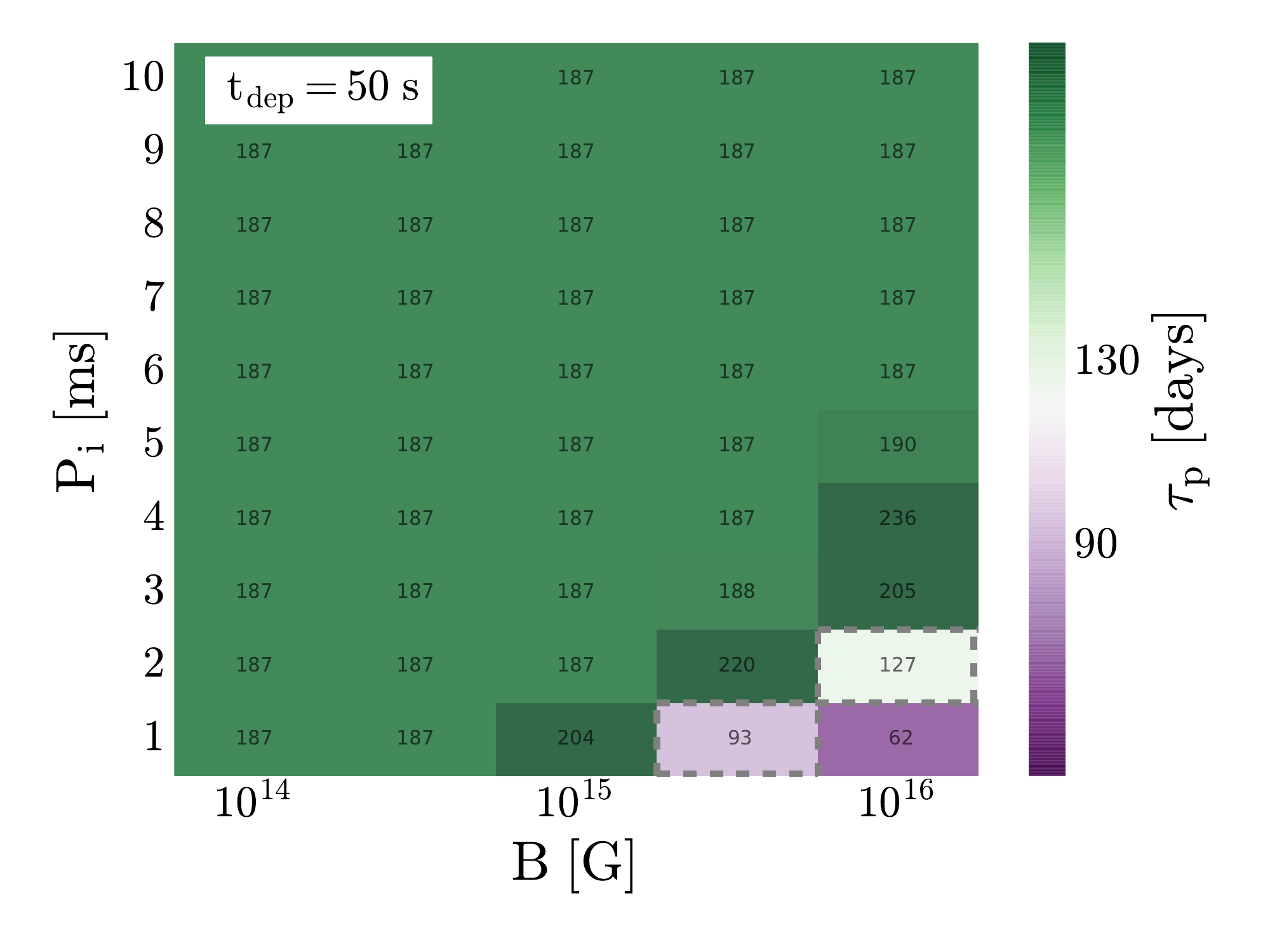}
\caption{Same as \Fig{tdepinf}, but here the magnetar 
deposits energy only during the initial 50 seconds. The range of 
magnetar parameters that transforms the weak explosion into a typical 
Type IIP light curve is now $\rm log(B)>10^{15}$ G and $\rm P_i <3$ 
ms. \lFig{tdep50}}
\end{figure}

Although the evidence is growing to connect the nature of the 
most energetic explosions (e.g., long duration gamma-ray bursts 
and SLSN-I) to a rotational energy source, imagining a magnetar 
being responsible for some ordinary Type IIP explosions is not 
too far fetched. So far we have detected only a few dozen 
magnetars in the Galaxy, and some of them are known to reside 
in Type II supernova remnants (based on their large ejecta mass) 
that seem to indicate a canonical explosion energy of 
$\sim10^{51}$ ergs \citep{Vin06}. One can also roughly 
approximate the vacuum spin-down periods for the 5 magnetars 
listed in the the McGill Online Magnetar Catalogue\footnote{http://www.physics.mcgill.ca/$\sim$pulsar/magnetar/main.html} 
\citep{Ola14}, that have a clear association with a supernova 
remnant. Using the measured surface dipole field strengths, 
assuming a breaking index of $\rm n=3$, and $\rm P_i=4$ ms, 
would bring the vacuum spin-down periods at the estimated 
ages (as $\rm P_i(1+t_{age}/t_m)^{1/2}$) to of order
the measured periods. This of course ignores known factors of 
the evolving magnetic field strength, both potentially on short 
time scales due to cooling, convection, and dynamo effects, and 
on long time scales due to non-MHD dissiapation as in 
\citet{Vig13,Gul15}. Nevertheless this is suggestive that 
magnetars might be somehow connected to some ordinary Type 
IIP supernovae.

Given the presupernova conditions for a typical solar 
metallicity RSG progenitor from stellar evolution calculations 
with dynamo processes, it is not straightforward to obtain 
proto-neutron star magnetic fields greater than $10^{15}$ G and 
initial spins of just few ms \citep{Heg05}. Magnetic torques 
during the evolution result in a slower spinning iron core, and 
without invoking magneto-rotational instabilities, or some other 
magnetic amplification processes, the  flux compression alone 
is not sufficient for a strong enough field. However, we note 
that the general problem of understanding the effective angular 
momentum transport in stellar interiors is still very much an 
open question. Until the existing theories are tested against 
asteroseismological measurements, which are the only way of 
probing the internal rotation profiles of evolved stars. The 
existing seismological data, though only available for much lower 
mass stars at the moment, already challenge our current 
understanding of angular momentum transport in redgiants 
\citep[e.g.,][]{Deh15,Tay13}.

In some ways it is not surprising that magnetar models can fit 
nearly all kinds of explosion light curves, including some regular 
Type IIP, when a big fraction of the explosion energy reservoir is 
replaced with a simple model that allows us to conveniently 
control its budget (initial spin) and injection rate (constant 
dipole magnetic field strength). This of course could be a fine-tuning, 
or interesting evidence that the supernova mehcanism is 
connected to rotation and magnetic fields.

\section*{Acknowledgments}

We thank Stan Woosley, Chris Kochanek, Laura Lopez, Katie Auchetl, Ondrej Pejcha and John Beacom for helpful discussions and comments. We also thank Alex Heger for his contributions in developing the \texttt{KEPLER} code. TS is partly supported by NSF (PHY-1404311) to John Beacom.


\end{document}